\documentclass[sigconf]{acmart}

\usepackage{listings}
\usepackage{booktabs} % For formal tables
\usepackage{graphicx}
\usepackage{caption}
\usepackage{subcaption}
\usepackage{tabulary}
\usepackage{rotating}
\usepackage{array}
\usepackage{multicol}
\usepackage{multirow}
\usepackage{listings}

\setcopyright{rightsretained}
\newcommand{\reg}{\textsuperscript{\textregistered{}}}

\acmDOI{https://doi.org/10.1145/3148173.3148189}

\acmISBN{978-1-4503-5565-0/17/11}

\acmConference[LLVM-HPC'17: ]{Fourth Workshop on the LLVM Compiler Infrastructure in HPC}{November 12--17, 2017}{Denver, CO, USA} 
\acmYear{2017}
\copyrightyear{2017}
\acmPrice{15.00}
\setcopyright{acmcopyright}

\begin{document}
\title{Implementing implicit OpenMP data sharing on GPUs}

\author{Gheorghe-Teodor Bercea}
\affiliation{%
  \institution{IBM TJ Watson Research Center}
  \streetaddress{1101 Kitchawan Rd.}
  \city{Yorktown Heights}
  \state{NY}
  \postcode{10598}
  \country{USA}}
\email{gheorghe-teod.bercea@ibm.com}

\author{Carlo Bertolli}
\affiliation{%
  \institution{IBM TJ Watson Research Center}
  \streetaddress{1101 Kitchawan Rd.}
  \city{Yorktown Heights}
  \state{NY}
  \postcode{10598}
  \country{USA}}
\email{cbertol@us.ibm.com}

\author{Arpith C. Jacob}
\affiliation{%
  \institution{IBM TJ Watson Research Center}
  \streetaddress{1101 Kitchawan Rd.}
  \city{Yorktown Heights}
  \state{NY}
  \postcode{10598}
  \country{USA}}
\email{acjacob@us.ibm.com}

\author{Alexandre Eichenberger}
\affiliation{%
  \institution{IBM TJ Watson Research Center}
  \streetaddress{1101 Kitchawan Rd.}
  \city{Yorktown Heights}
  \state{NY}
  \postcode{10598}
  \country{USA}}
\email{alexe@us.ibm.com}

\author{Alexey Bataev}
\affiliation{%
  \institution{IBM TJ Watson Research Center}
  \streetaddress{1101 Kitchawan Rd.}
  \city{Yorktown Heights}
  \state{NY}
  \postcode{10598}
  \country{USA}}
\email{alexey.bataev@ibm.com}

\author{Georgios Rokos}
\affiliation{%
  \institution{IBM TJ Watson Research Center}
  \streetaddress{1101 Kitchawan Rd.}
  \city{Yorktown Heights}
  \state{NY}
  \postcode{10598}
  \country{USA}}
\email{grokos@us.ibm.com}

\author{Hyojin Sung}
\affiliation{%
  \institution{IBM TJ Watson Research Center}
  \streetaddress{1101 Kitchawan Rd.}
  \city{Yorktown Heights}
  \state{NY}
  \postcode{10598}
  \country{USA}}
\email{hsung@us.ibm.com}

\author{Tong Chen}
\affiliation{%
  \institution{IBM TJ Watson Research Center}
  \streetaddress{1101 Kitchawan Rd.}
  \city{Yorktown Heights}
  \state{NY}
  \postcode{10598}
  \country{USA}}
\email{chentong@us.ibm.com}

\author{Kevin O'Brien}
\affiliation{%
  \institution{IBM TJ Watson Research Center}
  \streetaddress{1101 Kitchawan Rd.}
  \city{Yorktown Heights}
  \state{NY}
  \postcode{10598}
  \country{USA}}
\email{caomhin@us.ibm.com}

\renewcommand{\shortauthors}{Bercea G. et al.}

\begin{abstract}
OpenMP is a shared memory programming model which supports the offloading of target regions to accelerators such as NVIDIA GPUs. The implementation in Clang/LLVM aims to deliver a generic GPU compilation toolchain that supports both the native CUDA C/C++ and the OpenMP device offloading models. There are situations where the semantics of OpenMP and those of CUDA diverge. One such example is the policy for implicitly handling local variables. In CUDA, local variables are implicitly mapped to thread local memory and thus become private to a CUDA thread. In OpenMP, due to semantics that allow the nesting of regions executed by different numbers of threads, variables need to be implicitly \emph{shared} among the threads of a contention group.

In this paper we introduce a re-design of the OpenMP device data sharing infrastructure that is responsible for the implicit sharing of local variables in the Clang/LLVM toolchain. We introduce a new data sharing infrastructure that lowers implicitly shared variables to the shared memory of the GPU.

We measure the amount of shared memory used by our scheme in cases that involve scalar variables and statically allocated arrays. The evaluation is carried out by offloading to K40 and P100 NVIDIA GPUs. For scalar variables the pressure on shared memory is relatively low, under 26\% of shared memory utilization for the K40, and does not negatively impact occupancy. The limiting occupancy factor in that case is register pressure. The data sharing scheme offers the users a simple memory model for controlling the implicit allocation of device shared memory.

%In this paper we introduce a re-design of the OpenMP device data sharing infrastructure that is responsible for the implicit sharing of local variables in the Clang/LLVM toolchain. In previous work we proposed and implemented a more general GPU compilation toolchain that supports both the native CUDA C/C++ and the OpenMP device offloading models. Having one toolchain for both models promotes code reuse and enables the OpenMP implementation to be less susceptible to platform specific changes. However, there are situations where the semantics of OpenMP and those of CUDA diverge. One such example is the policy for implicitly handling local variables. In CUDA, local variables are implicitly mapped to thread local memory and thus become private to a CUDA thread. Due to OpenMP semantics that allow the nesting of regions executed by different numbers of threads, variables need to be implicitly visible to one ore more other threads within the same contention group. We introduce a new data sharing infrastructure that lowers implicitly shared variables to the shared memory of the GPU. We show that for many use cases the pressure on shared memory is relatively low, particularly for configurations likely to be encountered in practice. The evaluation is carried out by offloading to K40 and P100 NVIDIA GPUs.
\end{abstract}




\maketitle

\lstset{language=C++,
                basicstyle=\ttfamily,
                keywordstyle=\color{blue}\ttfamily,
                stringstyle=\color{red}\ttfamily,
                commentstyle=\color{green}\ttfamily,
                morecomment=[l][\color{magenta}]{\#}
}

\section{Introduction}

The increasingly wide adoption of the OpenMP\footnote{The following symbols used throughout the paper are registered trademarks: OpenMP\reg, NVIDIA\reg, IBM\reg,CUDA\reg} programming model in conjunction with heterogeneous architectures  has led to support for OpenMP device offloading being integrated into all the major compilers with Clang/LLVM at the forefront of this development effort.

Although, in this paper, OpenMP device offloading only targets NVIDIA GPUs, the discussion aims to be extensible to any existing toolchain that can become a target of the OpenMP device offloading model. From an OpenMP perspective, this paper is focused on the redesign of the implicit sharing of variables across threads which is one of the the most challenging aspects of our OpenMP support with ramifications touching both correctness and performance.

The principle that has guided the development of the OpenMP device offloading support for GPUs was to reuse, as much as possible, the existing OpenMP  support for the host as well as the existing device specific support in the LLVM backend. The resulting implementation is a generalization of the existing CUDA toolchain which has been extended to support the compilation of OpenMP target regions. In doing so, we aim to have our implementation automatically build on any future changes to NVIDIA devices and the OpenMP standard.

The device specific backend, i.e. the NVPTX backend of LLVM, has, so far, not been involved in any of the changes to generalize the CUDA toolchain to support OpenMP target regions. The NVPTX backend contains several assumptions strongly aligned with the native CUDA C/C++ programming model and less so with OpenMP. In this paper we will address one such assumption that is strongly connected with the sharing of variables among OpenMP threads: the implicit lowering of local variables to thread-local memory.

\subsection{Contributions}

In this paper we make the following contributions:
\begin{itemize}
\item We introduce a new implicit allocation policy in the NVPTX backend of LLVM to lower implicitly shared OpenMP variables to the shared memory of the device.
\item We introduce a redesign of the data sharing scheme on top of Clang trunk and libomptarget adapted to the changes to the LLVM NVPTX backend mentioned in the previous contribution.
\item We evaluate the impact on the shared memory usage of the new data sharing scheme to stress test the limits of the applicability of the scheme on both NVIDIA K40 and P100 GPUs.
\end{itemize}

\subsection{Background}
The existing support for OpenMP target regions is built on top of the host implementation of OpenMP and is confined, almost exclusively, to the Clang frontend code generation module. The most recent code generation scheme for OpenMP target regions is detailed in~\cite{arpith} and is based on previous work~\cite{llvm4hpc1, llvm4hpc2, llvm4hpc3} covering data-parallel cases~\cite{pmbs1, pmbs2} as well as nested parallelism~\cite{arpith}.

\subsection{Overview}

In Section~\ref{sec:rw} we give an overview of previous work that involves the Clang/LLVM OpenMP device offloading toolchain. Section~\ref{sec:openmp} contains a set of examples in which implicit data sharing is required by the OpenMP standard. Section~\ref{sec:codegen} details the code generation practices used in the most recent version of OpenMP device offloading. In Section~\ref{sec:implementation} we introduce changes to Clang, LLVM and libomptarget. The evaluation of the data sharing scheme is performed in Section~\ref{sec:evaluation}.

\section{Related Work}
\label{sec:rw}
Nested parallelism and data sharing across CUDA threads on NVIDIA GPUs
are the main subjects of~\cite{cuda-np}. The paper introduces a CUDA
language extension for nested parallelism, where CUDA kernels can
contain OpenMP-like pragmas that mark when a loop can be executed in
parallel.  When executing, a set of master threads (e.g. one per warp)
are active in the kernel in all regions outside of the ones marked as
nested parallel; slave threads (e.g. remaining threads in each warp)
are only activated when executing in the nested parallel
regions. Under this scheme, a data sharing problem is defined when
slave threads need to access data declared in the original CUDA
program as local variables owned by their master thread. Unlike our
contribution, this paper describes a source-to-source compiler.

The authors identify two scenarios under which data sharing
is required and present compiler solutions to these. First,
scalar variables declared as locals can be shared either using shift
instructions, if available, or shared memory. To enable sharing from a
sequential to a parallel region, the compiler inserts special sharing
function calls that mask the actual implementation (shifting or shared
memory). Three cases of sharing from a parallel to a sequential region
are identified: reductions and scans, which can be implemented using
shifts or shared memory depending on the parallelism scheme adopted; a
special case that corresponds to the {\it conditional} modifier of
lastprivate in OpenMP~\cite{omp5-tr4}. This is implemented as a
reduction and through pre-initialization of the base variable.

Second, static arrays declared as locals can be mapped using three
strategies, namely using (i) global memory, (ii) shared memory, or by
(iii) partitioning into multiple smaller local memory arrays. Local
array partitioning (iii) is used when the compiler can prove that each
slave will only access its assigned array partition. If that cannot be
proved, then the compiler selects shared memory (ii) if the size of
the local array is smaller than a pre-determined constant. If shared
memory cannot be used, then the compiler fallbacks into a global
memory mapping (i).

The solution presented in this paper maps all local variables onto
shared memory and we describe future work in the direction of falling
back into global memory.  Falling back will be based on similar
conclusion as those mentioned in~\cite{cuda-np}. This paper shows an
implementation of data sharing within the LLVM NVPTX backend for the
OpenMP language. It relies on high-level information from Clang for
optimization purposes.

Our contribution and~\cite{cuda-np} rely on the presence of implicit
and compiler-inserted barriers on GPUs to guarantee a consistent view
of data stored in shared and global memory amongst cooperating
threads. This is also the case of~\cite{openmp-ti}, which describes an
OpenMP implementation of a special-purpose DSP accelerator. The paper
describes a software-based memory coherence mechanism for shared
variables - no data sharing mechanism is required in this
implementation for private data. This is based on introducing memory
consistency operations (e.g. write-back) at appropriate OpenMP flush
points, when necessary.

\section{Implicit sharing of variables between OpenMP threads}
\label{sec:openmp}

OpenMP supports the nesting of code regions executed by different numbers of threads. There are numerous use cases in practice where variables need to be implicitly shared among several or all the threads of a team.

\subsection{OpenMP directives}

Throughout the paper we will use \emph{OpenMP code block} to mean the user code in between two consecutive OpenMP directives which potentially alter the number of threads. Within the same execution unit launched on the device (i.e. the same kernel) the number of active threads may vary depending on which OpenMP code block is being executed: the code in between a target and a parallel may be executed by one thread while the code inside the parallel is executed by all threads. OpenMP semantics allows for an OpenMP code block to be single-threaded, fully-threaded or be executed by a user-defined number of threads.

\subsubsection{\texttt{target} regions}
A target region is a code block associated with a \texttt{\#pragma omp target} directive. Target regions are compiled for the device specified via the \texttt{-fopenmp-target=<target\_triple>} compiler flag. In the case of NVIDIA K10 and P100 GPUs the target triple is given by \texttt{nvptx64-nvidia-cuda}. A target region that contains no other OpenMP directives is executed by a single thread.

A target region may contain a \texttt{teams} directive in which case, the two directives must be closely nested i.e. there exists no OpenMP code block in between the two directives. Apart from this case, OpenMP code blocks can occur between any other two consecutive directives that affect parallel execution: \texttt{teams}, \texttt{distribute}, \texttt{parallel}, \texttt{for} and \texttt{simd}. 

\subsubsection{\texttt{parallel} regions}
A code block associated with a \texttt{\#pragma omp parallel} directive is executed by all available threads of a contention group unless the user specifies a custom number of threads via the \texttt{thread\_limit} clause. All threads execute the same parallel region body unless a worksharing construct such as a \texttt{\#pragma omp for} is encountered. The team of threads executing the parallel region that encounter the worksharing construct cooperatively execute its associated code block. OpenMP allows for parallel regions to be nested further increasing the number of possible regions executed by different threads.

\subsection{Implicitly shared variables}

The \texttt{target} and \texttt{parallel} directives are sufficient for constructing an execution unit within which the number of active threads varies. Example is shown in Figure~\ref{fig:target-parallel}. Figure~\ref{fig:target-parallel1} contains a single non-empty OpenMP code block where a local variable \texttt{c} is declared. Variable \texttt{c} is local to each thread and is implicitly private to each thread.

\begin{figure}[t!]
\centering
\begin{subfigure}[h]{0.4\textwidth}
\begin{lstlisting}
#pragma omp target
{
  #pragma omp parallel for
  for (int i = 0 ; i < N ; i++) {
    int c = 1;
    // read/write c
  }
}
\end{lstlisting}
\caption{Device-offloaded target region with innermost declaration and use of variable \texttt{c}.}
\label{fig:target-parallel1}
\end{subfigure}
\begin{subfigure}[b]{0.4\textwidth}
\begin{lstlisting}
#pragma omp target
{
  // Code block 1
  int c = 1;
  #pragma omp parallel for
  for (int i = 0 ; i < N ; i++) {
    // Code block 2
    // read c
  }
}
\end{lstlisting}
\caption{Device-offloaded target region with declaration of variable \texttt{c} in the target-only region.}
\label{fig:target-parallel2}
\end{subfigure}
\caption{Example OpenMP programs.}
\label{fig:target-parallel}
\end{figure}

In Figure~\ref{fig:target-parallel2} the code block in between the \texttt{target} and the \texttt{parallel} directives is not empty. OpenMP semantics for target regions require that this code block be executed by a single thread. In a real-world use case the code in this code block may contain side effects which would lead to incorrect multithreaded execution.

In this latter example, variable \texttt{c} is local to the one thread executing the target-only region but has to be available to all threads executing the parallel construct. In the absence of clauses that would alter implicit sharing, it is essential to the correctness of the program that variable \texttt{c} be implicitly shared by the master thread with all the threads executing the parallel region.

In general, according to the OpenMP 4.5 specification, implicitly shared variables are defined as variables referenced inside a given construct and do not have predetermined data-sharing attributes, and are not listed in a data-sharing attribute clause on the construct. There are several cases:
\begin{itemize}
\item When these variables are inside a \texttt{teams}, \texttt{parallel} or task generating construct, the data-sharing attributes are determined by the \texttt{default} clause if present. In a parallel construct, if no default clause is present, these variables are shared.
\item For constructs other than task generating constructs or target constructs, if no default clause is present, these variables reference the variables with the same names that exist in the enclosing context.
\item In a target construct, variables that are not mapped after applying data-mapping attribute rules are firstprivate. In an orphaned task generating construct, if no default clause is present, formal arguments passed by reference are firstprivate.
\item In a task generating construct, if no default clause is present, a variable for which the data-sharing attribute is not determined by the rules above and that in the enclosing context is determined to be shared by all implicit tasks bound to the current team is shared.
\item In a task generating construct, if no default clause is present, a variable for which the data-sharing attribute is not determined by the rules above is firstprivate.
\end{itemize}
In this paper we will focus on the simplest example which contains a \texttt{parallel} construct nested inside a \texttt{target} region.

\section{OpenMP code generation in Clang}
\label{sec:codegen}

\begin{figure}[ht!]
\centering
\begin{subfigure}[b]{0.4\textwidth}
\begin{lstlisting}
entry:
  %1 = icmp ult i32 %nvptx_tid,
    %thread_limit
  br i1 %1, label %worker,
    label %mastercheck

worker:
  call void @WORKER()
  br label %exit

mastercheck:
  %2 = icmp eq i32 %nvptx_tid,
    %master_tid
  br i1 %2, label %master, label %exit

master:
  call void @__kmpc_kernel_init(
    i32 %thread_limit)
  call void @__kmpc_kernel_prepare_parallel()
  ; Code block 1 emitted here
  call void @llvm.nvvm.barrier0()
  call void @llvm.nvvm.barrier0()
  br label %exit

exit:
  ret void
\end{lstlisting}
\end{subfigure}
\caption{Generated code for OpenMP master region.}
\label{fig:codegen-master}
\end{figure}

\begin{figure}[ht!]
\centering
\begin{subfigure}[b]{0.4\textwidth}
\begin{lstlisting}
entry:
  br label %await.work

await.work:
  ; master stops execution
  call void @llvm.nvvm.barrier0()
  call i1 @__kmpc_kernel_parallel(
    i8** %work_fn)
  br i1 %terminate, label %exit,
    label %select.workers

select.workers:
  %3 = load i8, i8* %exec_status
  %is_active = icmp ne i8 %3, 0
  br i1 %is_active,
    label %execute.parallel,
    label %barrier.parallel

execute.parallel:
  ; function contains Region 2
  call void @OUTLINE_PARALLEL(
    i16 0, i32 %master_tid, [...]) ; args
  br label %terminate.parallel

terminate.parallel:
  call void @__kmpc_kernel_end_parallel()
  br label %barrier.parallel

barrier.parallel:
  ; master resumes execution
  call void @llvm.nvvm.barrier0()
  br label %await.work

exit:
  ret void
\end{lstlisting}
\end{subfigure}
\caption{Generated code for OpenMP worker region.}
\label{fig:codegen-worker}
\end{figure}

In our latest publication on the Clang code generation~\cite{arpith} we introduce a software implementation of the fork-join model for the GPU. This new scheme is based on dynamic assignment of pointers to outlined functions (i.e. the actual workloads) to a pool of parallel threads. We also refer to this scheme as \emph{dynamic work allocation}. This scheme models the flexibility of the fork-join model on which the OpenMP programming language is based.

Due to the fact that we target NVIDIA GPUs, in this section and in the remainder of the paper we will use the OpenMP terminology of \emph{teams} and \emph{threads} as a one-to-one mapping to the CUDA model concepts of \emph{threadblocks} and \emph{CUDA threads} respectively.

The code generation scheme manages the threads within a contention group and ensures that OpenMP semantics are respected. OpenMP supports the nesting of several parallel construct types. In this paper we focus on the simplest example for which the new code generation is employed: the nesting of a parallel region inside a sequential region. An example of this use case is shown in Figure~\ref{fig:target-parallel2} where code block 1 is executed sequentially whilst code block 2 is parallel.

The sequential region must be executed by only one thread, due to potential side effects in the user code. We also want to support cases in which functions called from within this region may contain other OpenMP parallel constructs. The latter aspect is handled by the dynamic work allocation scheme in which the \emph{master} thread allocates any outstanding parallel workloads to be cooperatively executed by a group of \emph{worker} threads. Work allocation entails the passing of a pointer to an outlined function containing an individual parallel workload.

Target regions are compiled down to kernels launched on the GPU with a predetermined\footnote{The number of teams and threads per team are determined by the user or, if not, by the runtime.} number of teams and threads per team. Within a team, the threads are all uniformly launched. Within this uniform pool of threads, the code generation scheme assigns all threads in order of the thread identifier to the worker pool except for the last 32 threads. The last 32 threads are reserved for the master thread region. The last 32 threads correspond to a full CUDA warp of threads~\footnote{For a detailed explanation of this choice please refer to~\cite{arpith}.}. From within the 32 threads we select the first as the master and deactivate the remaining 31. The LLVM-IR code generated for this scheme is shown in Figure~\ref{fig:codegen-master}. Once the master thread is isolated from the rest of the threads, it executes any sequential region and assigns the workload for all the worker threads to execute.

Named barriers are used to control the execution of master and worker threads. Whilst the master executes, the workers wait at a barrier and vice-versa. The code for the workers is shown in Figure~\ref{fig:codegen-worker}. The outlined parallel region is a function pointer passed by the master to the worker threads.

\section{Implicit sharing of variables in Clang/LLVM}
\label{sec:implementation}

In the previous section we outlined the code generation scheme employed by Clang. We consider it the baseline implementation on top of which we introduce the new data sharing infrastructure.

Apart from any device specific challenges, the new data sharing infrastructure needs to deal with additional issues introduced by the previous code generation scheme: (1) the data sharing infrastructure requires the sharing of variables across different functions, i.e. from the master function to the worker function, (2) the worker function must be able to handle multiple outlined workloads and (3) each outlined workload may need to access a unique combination of implicitly shared variables.

On the device specific side the challenges are: (1) due to no communication between Clang and LLVM outside the code Clang generates, implicitly shared variables must be detected in the LLVM backend, (2) variables which need to be implicitly shared must be allocated in a shareable address space of the GPU device.

In this section we will cover changes to three different packages: Clang, LLVM and libomptarget. The main data sharing infrastructure is discussed in relation to the master-worker data sharing outlined in Section~\ref{sec:codegen}.

\subsection{Clang code generation}
\label{sec:clang}

To share a value across functions, we rely on the runtime to set up an array of references to the shared values. In the following section we will describe how the runtime manages this list. In this section we will focus on the changes to Clang code generation. Throughout this section we assume that the address of any shared variable can be shared among threads.

In Figure~\ref{fig:target-parallel2} we have a simple data sharing example which we will use to describe the changes to the Clang code generation scheme.

Both the master and any or all the worker threads may require read and write access to a shared variable. Making sure that the most up to date value is used, we require the sharing of a reference to this value instead. The master and worker threads can then follow this reference every time access to the variable is required for either reading or writing. Due to the way the code generation scheme in Section~\ref{sec:codegen} works, there are no race conditions between master- and worker-thread accesses. Race conditions across workers are handled at user level.

We first create a reference to variable \texttt{c} by invoking the appropriate \texttt{alloca} instruction. We also create the pointer to the list of shared references, \texttt{shared\_args}:
\begin{lstlisting}
define void @KERNEL(
    i32* dereferenceable(4) %c){
entry:
  %c.addr = alloca i32
  store i32 %c, i32* %c.addr
  %shared_args = alloca i8**
\end{lstlisting}

In the master only region, we invoke the runtime function \texttt{\_\_kmpc\_kernel\_prepare\_parallel} augmented with the reference to the list of shared argument references along with the number of shared variables:
\begin{lstlisting}
  call void @__kmpc_kernel_prepare_parallel(...,
    i8*** %shared_args, i32 1)
\end{lstlisting}

The runtime will return a reference to a list of the desired length which the master begins to initialize with the references to the shared variables. In our example there exists only one such value that requires initialization and the following code is emitted following the runtime call above:
\begin{lstlisting}
  %17 = load i8**, i8*** %shared_args
  %18 = getelementptr inbounds i8*,
    i8** %17, i64 0
  %19 = bitcast i32* %c.addr to i8*
  store i8* %19, i8** %18
\end{lstlisting}

Once the value is set, the master-only region can access the shared variable via its reference. Any updates to the variable in the master region will therefore be visible to any worker thread that follows the same reference.

The worker function requires minor changes to handle the passing in of the \texttt{shared\_args} list. The worker function interacts with the runtime via the \texttt{\_\_kmpc\_kernel\_parallel} function. The function has been extended to support this. Following this call, each worker thread obtains a handle on the list of shared variables.
\begin{lstlisting}
  call i1 @__kmpc_kernel_parallel(
    i8** %work_fn,
    i8*** %shared_args)
\end{lstlisting}

The list of arguments is potentially unique to every outlined region and each worker needs to know the way in which the outlined function is called. We construct a special function called a \texttt{wrapper} function which passes the arguments to the parallel outlined region including any shared arguments. Each worker, instead of calling the outlined parallel region directly, will call the wrapper instead. The wrapper arguments include the list of shared arguments:
\begin{lstlisting}
  %5 = load i8**, i8*** %shared_args
  call void @WRAPPER(..., i8** %5)
\end{lstlisting}

The wrapper function is shown in Figure~\ref{fig:codegen-wrapper}. The wrapper function controls the order of the parameters by passing them in the same order they appear in the list of shared arguments.

\begin{figure}[ht!]
\centering
\begin{subfigure}[b]{0.4\textwidth}
\begin{lstlisting}
define void @WRAPPER(..., i8**){
entry:
  %c.addr = alloca i32*
  %.addr2 = alloca i8**
  store i8** %2, i8*** %.addr2,
  bra label %next

next:
  %3 = load i8**, i8*** %.addr2
  %4 = getelementptr inbounds i8*,
    i8** %3, i64 0
  %5 = bitcast i8** %4 to i32**
  %6 = load i32*, i32** %5
  call void @OUTLINE_PARALLEL(
    i32* null, i32* null, i32* %6)
  bra label %exit

exit:
  ret void
}
\end{lstlisting}
\end{subfigure}
\caption{Generated code for OpenMP wrapper function which passes any arguments which come from data sharing to the outlined parallel function.}
\label{fig:codegen-wrapper}
\end{figure}

\subsection{libomptarget: support list of references to shared variables}
\label{sec:libomptgt}

The changes to the runtime include changes to the interface to accommodate the passing of the list of references to shared variables and its allocation.

The list of references to shared variables consists of a statically preallocated list in the shared memory of the device and is 20 entries in length. On the K40 and the P100 NVIDIA GPUs this leads to a shared memory footprint of 160 bytes per threadblock (or OpenMP team). Note that this list only needs to hold the references to the shared arguments so it only needs to handle the number of shared entities regardless of whether they are scalars or statically declared arrays.

The length of the list has been empirically chosen based on our limited application experience and is a conservative figure. With the increases in shared memory on newer GPU models such as the NVIDIA P100, the size of the preallocated list can be increased.

When the size of this list is insufficient, the back-up scheme is to dynamically allocate a list of variables using the \texttt{malloc} function. The list will therefore be allocated in the global memory of the device at the beginning of the parallel region and deallocated at the end. This back-up scheme is designed as a correctness safety-net. The shared memory implementation on the other hand, is designed to deliver lower latency accesses. Experiments show that the different in performance between the two schemes can be as large as an order of magnitude.

\subsection{Generalizing the LLVM NVPTX backend}
\label{sec:llvm}

The design principle guiding the development of the OpenMP device offloading toolchain for NVIDIA GPUs was to generalize the functionality already exposed in the CUDA toolchain of Clang/LLVM. There are several reasons for advocating for a more general toolchain: (1) code reuse of accelerator specific parts of the code base and (2) keeping up with any NVIDIA specific architectural changes. Tools like NVPTX will always be kept up to date with the latest CUDA releases so having that as part of the toolchain increases the long term maintainability of any OpenMP device offloading toolchain and reduces code duplication.

The new scheme in Clang code generation for implicitly shared variables described in Sections~\ref{sec:clang} and~\ref{sec:libomptgt} relies on the addresses of the variables being shareable among threads. On NVIDIA GPUs only variables in shared or global memory can have their addresses shared across the threads of an OpenMP team.

The LLVM NVPTX backend lowers all the locally allocated variables via the \texttt{alloca} LLVM-IR instruction to a thread local memory stack which is emitted at the level of the PTX code. Lowering variables to thread local memory is in line with the CUDA programming model and it is enough to satisfy its requirements. For OpenMP a more generic allocation policy of local variables is required. In the remainder of the section we discuss the lowering of local variables to the shared memory of the device for the cases required by OpenMP.

\subsubsection{Shared memory stack}
To allow for the lowering of variables to shared memory the prologue of the output PTX kernel function is augmented with a stack allocated in the shared memory of the device:

\begin{lstlisting}
.local  .align 8 .b8 __local_depot[10]
.shared .align 8 .b8 __shared_depot[10]
\end{lstlisting}

The shared memory depot is of the same size as the local memory depot. This allows us to reuse the local offsets within the shared stack as well. In general this is wasteful for shared memory and we aim to optimize this in future work. The evaluation of the amount of shared memory is shown in Section~\ref{sec:evaluation}.

The shared stack implementation requires a shared stack pointer. We create a special register similar to the local stack pointer which we add to the prologue of the function:

\begin{lstlisting}
mov.u64         %SPL, __local_depot
mov.u64         %SPSH, __shared_depot
cvta.local.u64  %SP, %SPL
cvta.shared.u64 %SP, %SPSH
\end{lstlisting}

The next step after creating the shared memory stack is to use the shared stack pointer for those cases where a variable is shared under OpenMP semantics. The LLVM intermediate representation does not attach any specific memory information to the allocation instruction \texttt{alloca}. An LLVM backend may choose to lower the \texttt{alloca} instruction to a any available memory types available on the device.

\subsubsection{Detecting shared variables}
Until the addition of the shared memory stack, the NVPTX backend had only one option: implicitly mapping any \texttt{alloca} instruction to the thread's local memory. With the addition of the shared memory stack, the NVPTX backend  needs to now choose which variables to lower to shared memory and which to lower to local memory. If we consider local memory allocation to be the default behavior,  the NVPTX backend needs to contain a way to detect which variables should be lowered to the device shared memory.

In this initial implementation, the detection of shared variables is done on the basis of their address being taken. In the code generation strategy described in Section~\ref{sec:clang} shared variables have their address taken and stored in the array of references to shared variables. Every shared variable is therefore guaranteed to have its address taken at least once.\

Checking whether the address is being taken is straightforward. We iterate through the uses of a given \texttt{alloca}. Whenever a reference to the allocated value is stored we assume that the variable is shared:

\begin{lstlisting}
    // Check if Ptr or an alias to it is
    //the destination of the store
    auto SI = dyn_cast<StoreInst>(Use);
    if (SI)
      for (auto Alias: PointerAliases)
        if (SI->getValueOperand() == Alias)
          return true; // address is taken
\end{lstlisting}

A list of \emph{aliases} needs to be maintained in case the address of an alias of the original value is taken. For example, when the original value is put through a bit cast instruction.

The detection of shared variables and their lowering to shared memory needs to happen at the same time local variables are lowered to thread local memory in the \texttt{LowerAlloca} pass of NVPTX. We augment the pass with the means to detect whether the address of a particular variable is taken. This includes the taking of an address of any aliases of the original value returned by the \texttt{alloca}.

In the cases when the compiler is invoked with the \texttt{-O0} flag, the \texttt{LowerAlloca} pass is not invoked. This means that an alternative pass needs to be create for this purpose. We call this new pass the \texttt{FunctionDataSharing} pass. This pass relies on the same detection strategy as the \texttt{LowerAlloca} pass.

\subsubsection{Lowering \texttt{alloca} instructions to device shared memory}
Following the detection stage we need to lower the variables to the shared memory of the device. This can be achieved by inserting the address space cast instruction twice: to cast the variable from the generic to the shared address space immediately followed by an address space cast from the shared to the generic address space. This will enable subsequent passes to link the original \texttt{alloca} with the usage of shared memory. This will also enable subsequent load and store instructions to bind to the shared version of the variable thus enabling the usage of specific instructions such as \texttt{ld.shared} and \texttt{st.shared}.

NVPTX contains several optimization passes over machine instructions. Address space casts are inserted just before the LLVM-IR code is translated into machine instructions. This ensures that the memory type information we inserted is preserved and passed down to the next abstraction level.

Machine instructions are the first level of abstraction that contains frame indices. This is where a frame index is mapped to the generic, local or the newly added shared stack pointer register, \texttt{VRShared}. The \texttt{VRShared} register is to the shared memory stack what the \texttt{VRFrameLocal} register is to the local memory stack.

The lowering of frame indices to the shared index is performed in a new pass we call the \texttt{LowerSharedFrameIndices} pass. This pass traverses the kernel function and for each frame index we encounter, we check if the index has been translated to the shared register already. If it has not been translated, we check if the result of the operation on that frame index is converted to shared memory using one of the instructions we inserted before. An example of the machine instruction pattern we need to identify is given by:
\begin{lstlisting}
%vreg25<def> = LEA_ADDRi64 <fi>, 0;
%vreg6<def> = cvta_to_shared_yes_64
    %vreg25<kill>;
\end{lstlisting}

If the pattern is detected, that frame index is replaced with the stack pointer of the shared frame:
\begin{lstlisting}
%vreg25<def> = LEA_ADDRi64 %VRShared, 0;
\end{lstlisting}

The frame index is then added to a list of already translated indices. This ensures that the shared frame index is propagated appropriately to all instructions that use it.

This \texttt{LowerSharedFrameIndices} pass needs to occur before the \texttt{StackColoring} pass to ensure correctness of the stack slot coloring algorithm. If not, the algorithm may lead to the same local stack slot being used by both a local and a shared variable. This leads to the generation of incorrect code. Since the stack slot coloring algorithm works on frame indices, the earlier lowering of frame indices to the shared memory register excludes those frame indices from being considered by the algorithm.

\section{Evaluation}
\label{sec:evaluation}

% When invoking a generic kernel, the worker function is also emitted. I believe this is one of the cases where we cannot tell if there will be a parallel region inside the target region in advance. It would be useful to know this so that we can just generate minimal code.

We exercise the generic code generation of OpenMP in the Clang compiler in combination with the implicit sharing of variables. In Section~\ref{sec:setup} we include the experimental setup, in Section~\ref{sec:results} we showcase the results and in Section~\ref{sec:discussion} we discuss the results.

\subsection{Experimental setup}
\label{sec:setup}

We test the new data sharing infrastructure on two NVIDIA GPUs, the K40 and the P100 GPU. They each feature a shared memory area on a per-SM basis. On the K40 shared memory and L1 cache share the same 64 KB of physical memory and can be configured in three different ways: 48KB L1 + 16KB shared memory, 32KB L1 + 32KB shared memory and 16KB L1 + 48KB shared memory. In the default configuration the K40 uses 16KB/SM as shared memory. The P100 GPU has 64 KB shared memory per SM, separate from the L1 cache.

The characteristics of the two GPUs that we are interested in are shown in Table~\ref{tab:gpu}. In all experiments on the K40 we use the default split between shared memory and L1 cache: 16KB of shared memory and 48 KB of L1 cache.

%--------------------------------------------------------------------------------
\begin{table}
\begin{center}
\begin{tabular}{| c | c | c |}
\hline

Feature & K40 GPU  & P100 GPU \\\cline{1-3}
%----------------------------------------
Concurrent blocks/SM & 16 & 32 \\\cline{1-3}
%----------------------------------------
32-bit Registers/SM & 65536 & 65536 \\\cline{1-3}
%----------------------------------------
Shared memory & 16KB/32KB/48KB & 64KB \\\cline{1-3}
%----------------------------------------
\end{tabular}
\end{center}
\caption{Comparison between K40 and P100 GPUs}
\label{tab:gpu}
\end{table}
%--------------------------------------------------------------------------------

We run two different test programs that exercise the implicit sharing of scalar and array variables. We evaluate the amount of shared memory in each case. We want to determine whether the shared memory is a limitation in these cases. The two programs are shown in Figures~\ref{fig:target-parallel-vars} and~\ref{fig:target-parallel-arrays}.

\begin{figure}[ht!]
\centering
\begin{subfigure}[b]{0.4\textwidth}
\begin{lstlisting}
#define WORKERS 96
#define SIZE TEAMS*WORKERS
void increment_array(int *a) {
  int c1 = 0;
  int c2 = 0;
  [...]
  int c7 = 0;
  #pragma omp target map(tofrom:a[:SIZE])
  #pragma omp teams num_teams(TEAMS)
thread_limit(WORKERS) 
  {
    c1 += 1;
    c1 += 2;
    [...]
    c7 += 7;
    #pragma omp parallel
    {
      a[omp_get_team_num()*WORKERS +
        omp_get_thread_num()] +=
         c1 + c2 + [...] + c7;
    }
  }
}
\end{lstlisting}
\end{subfigure}
\caption{Implicitly sharing 8 variables (including array \texttt{a}) between the \texttt{target} and \texttt{parallel} regions. }
\label{fig:target-parallel-vars}
\end{figure}

The program in Figure~\ref{fig:target-parallel-vars} exercises the data sharing infrastructure in isolation. The example only shows the case for sharing seven local variables and an array reference, requiring the sharing of eight values in total. To test the applicability of the scheme to more complex examples we test the scheme on up to 64 shared variables and note the impact on the amount of shared memory usage. 

\begin{figure}[ht!]
\centering
\begin{subfigure}[b]{0.4\textwidth}
\begin{lstlisting}
#define WORKERS 96
#define SIZE TEAMS*WORKERS
void increment_array(int *a) {
  #pragma omp target map(tofrom:a[:SIZE])
  #pragma omp team thread_limit(WORKERS)
num_teams(TEAMS)  
  {
    int d1[WORKERS];
    for(int i=0; i<WORKERS; i++)
      d1[i] = 10;
    #pragma omp parallel
    {
      a[omp_get_team_num()*WORKERS +
        omp_get_thread_num()] +=
         d1[omp_get_thread_num()];
    }
  }
}
\end{lstlisting}
\end{subfigure}
\caption{Implicitly sharing an array which has as many entries as there are worker threads.}
\label{fig:target-parallel-arrays}
\end{figure}

The program in Figure~\ref{fig:target-parallel-arrays} uses arrays instead of variables. This keeps the register usage low while increasing the pressure on shared memory. We test up to four arrays of equal size to the number of worker threads.

Each OpenMP team is mapped to a CUDA threadblock. This means that anything we share across the team can be mapped to shared memory directly. The number of teams covers the entire length of the output \texttt{a} array.

To be closer to setups used in practice, we fix the number of threads per team to 128. We use 96 worker threads in addition to the 32 threads used by the master warp. We use the \texttt{thread\_limit} clause to set the number of threads to 96 - the \texttt{thread\_limit} clause always sets the number of workers, the actual number of allocated threads is 32 higher than the thread limit to account for the master warp.

From our application experience, the most common choices for the number of threads per team is 128 or 256. Since shared memory is allocated on a team basis, we choose to test with 128 threads as this will lead to a higher number of teams and will be a stricter test for shared memory usage.

\subsection{Experimental results}
\label{sec:results}

\subsubsection{Kernel shared memory usage}
\label{sec:footprint}

We evaluate the shared memory usage of the programs for both K40 and P100 GPUs in Tables~\ref{tab:one} and~\ref{tab:four} respectively. The shared memory figures in these two tables are independent of the number of threads in a team.
%--------------------------------------------------------------------------------
\begin{table*}[ht!]
\begin{center}
\begin{tabular}{| c | c | c | c | c |}
\hline
  Number of  &  Shared stack  & Pre-alloc & Thread private state & Total \\
  variables  &  [Bytes]  & [Bytes] & size [Bytes] & [Bytes] \\\cline{1-5}
%----------------------------------------
1 & 24  & 160 & 49 & 233 \\\cline{1-5}
%----------------------------------------
2 & 32  & 160 & 49 & 241 \\\cline{1-5}
%----------------------------------------
4 & 48  & 160 & 49 & 257 \\\cline{1-5}
%----------------------------------------
8 & 80  & 160 & 49 & 289 \\\cline{1-5}
%----------------------------------------
16 & 144  & 160 & 49 & 353 \\\cline{1-5}
%----------------------------------------
32 & 272  & 160 & 49 & 481 \\\cline{1-5}
%----------------------------------------
64 & 528  & 160 & 49 & 737 \\\cline{1-5}
%----------------------------------------
\end{tabular}
\end{center}
\caption{Static shared memory analysis of individual variables on the K40 GPU and P100 GPUs.
%None of these numbers scale with the number of threads in a team.
}
\label{tab:one}
\end{table*}
%--------------------------------------------------------------------------------
%--------------------------------------------------------------------------------
\begin{table*}[ht!]
\begin{center}
\begin{tabular}{| c | c | c | c | c |}
\hline
  Number of  &  Shared stack  & Pre-alloc & Thread private state & Total \\
  arrays  &  [Bytes]  & [Bytes] & size [Bytes] & [Bytes] \\\cline{1-5}
%----------------------------------------
1 & 408  & 160 & 49 & 617 \\\cline{1-5}
%----------------------------------------
2 & 792  & 160 & 49 & 1001 \\\cline{1-5}
%----------------------------------------
3 & 1176  & 160 & 49 & 1385 \\\cline{1-5}
%----------------------------------------
4 & 1560  & 160 & 49 & 1769 \\\cline{1-5}
%----------------------------------------
\end{tabular}
\end{center}
\caption{Static shared memory analysis of local arrays on the K40 GPU and P100 GPUs.
%None of these numbers scale with the number of threads in a team.
}
\label{tab:four}
\end{table*}
%--------------------------------------------------------------------------------
%--------------------------------------------------------------------------------
\begin{table*}[ht!]
\begin{center}
\begin{tabular}{| c | c | c | c | c | c |}
\hline

  Number of  &  Static shared  & Dynamic global        &   &                    & Shared \\
   variables  &  memory per team  & memory per team& Registers & Teams/SM & memory per SM \\
                    &  [Bytes]  & [Bytes] &  & &  [Bytes] \\\cline{1-6}
%----------------------------------------
1 & 233 & 0 & 36 & 14 & 3262\\\cline{1-6}
%----------------------------------------
2 & 241 & 0 & 36 &  14 & 3374 \\\cline{1-6}
%----------------------------------------
4 & 257 & 0 & 36 &  14 & 3598 \\\cline{1-6}
%----------------------------------------
8 & 289 & 0 & 36 &  14 & 4046 \\\cline{1-6}
%----------------------------------------
16 & 353 & 0 & 40 &  12 & 4236 \\\cline{1-6}
%----------------------------------------
32 & 481 & 256 & 72 & 7 & 3367 \\\cline{1-6}
%----------------------------------------
64 & 737 & 512 & 136 & 3 & 2211  \\\cline{1-6}
%----------------------------------------
\end{tabular}
\end{center}
\caption{Shared memory footprint of implicitly shared variables on the K40 GPU in the common use case where the target region can be executed by an arbitrary number of teams, each team comprising 128 threads.}
\label{tab:two}
\end{table*}
%--------------------------------------------------------------------------------
%--------------------------------------------------------------------------------
\begin{table*}[ht!]
\begin{center}
\begin{tabular}{| c | c | c | c | c | c | c |}
\hline

  Number of  &  Static shared  & Dynamic global        &        &                    & Shared \\
   variables  &  memory per team  & memory per team& Registers & Teams/SM & memory per SM \\
                    &  [Bytes]  & [Bytes] &  &  &  [Bytes] \\\cline{1-6}
%----------------------------------------
1 & 233 & 0 & 31 & 16 & 3728\\\cline{1-6}
%----------------------------------------
2 & 241 & 0 & 31 &  16 & 3856 \\\cline{1-6}
%----------------------------------------
4 & 257 & 0 & 31 &  16 & 4112 \\\cline{1-6}
%----------------------------------------
8 & 289 & 0 & 31 & 16 & 4624 \\\cline{1-6}
%----------------------------------------
16 & 353 & 0 & 40 &  12 & 4236 \\\cline{1-6}
%----------------------------------------
32 & 481 & 256 & 71 &  7 & 3367 \\\cline{1-6}
%----------------------------------------
64 & 737 & 512 & 135 &  3 & 2211  \\\cline{1-6}
%----------------------------------------
\end{tabular}
\end{center}
\caption{Shared memory footprint of implicitly shared variables on the P100 GPU in the common use case where the target region can be executed by an arbitrary number of teams, each team comprising 128 threads.}
\label{tab:twotwo}
\end{table*}
%--------------------------------------------------------------------------------
%--------------------------------------------------------------------------------
\begin{table*}
\begin{center}
\begin{tabular}{| c | c | c | c | c | c | c | c |}
\hline

  Number of  & Array data &  Static shared   &  &       Potential              & Shared & Actual \\
   variables  & shared &  memory per team  & Registers & Teams/SM & memory per SM & Teams/SM \\
                    & [Bytes] &  [Bytes]  & & &  [Bytes]  & \\\cline{1-7}
%----------------------------------------
1 & 384 & 617 & 36 & 14 & 8638 & 14\\\cline{1-7}
%----------------------------------------
2 & 768 & 1001 & 36 & 14 & 14014 & 14 \\\cline{1-7}
%----------------------------------------
3 & 1152 & 1385 & 36 & 14 & 19390 & 11 \\\cline{1-7}
%----------------------------------------
4 & 1536 & 1769 & 36 &  14 & 24766 & 9 \\\cline{1-7}
%----------------------------------------
\end{tabular}
\end{center}
\caption{Shared memory footprint of implicitly shared arrays on the K40 GPU in the common use case where the target region can be executed by an arbitrary number of teams, each team comprising 128 threads. We assume a shared memory configured at 16KB.}
\label{tab:five}
\end{table*}
%--------------------------------------------------------------------------------
%--------------------------------------------------------------------------------
\begin{table*}
\begin{center}
\begin{tabular}{| c | c | c | c | c | c | c |}
\hline

  Number of  & Array data &  Static shared   &  &      Potential              & Shared & Actual \\
   variables  & shared &  memory per team  & Registers & Teams/SM & memory per SM & Teams/SM \\
                    & [Bytes] &  [Bytes]  & & &  [Bytes]  & \\\cline{1-7}
%----------------------------------------
1 & 384 & 617 & 30 & 17 & 10489 & 17\\\cline{1-7}
%----------------------------------------
2 & 768 & 1001 & 30 & 17 & 17017 & 17 \\\cline{1-7}
%----------------------------------------
3 & 1152 & 1385 & 30 & 17 & 23545 & 17 \\\cline{1-7}
%----------------------------------------
4 & 1536 & 1769 & 30 & 17 & 30073 & 17 \\\cline{1-7}
%----------------------------------------
\end{tabular}
\end{center}
\caption{Shared memory footprint of implicitly shared arrays on the P100 GPU in the common use case where the target region can be executed by an arbitrary number of teams, each team comprising 128 threads.}
\label{tab:six}
\end{table*}
%--------------------------------------------------------------------------------
%--------------------------------------------------------------------------------
\begin{table*}
\begin{center}
\begin{tabular}{| c | c | c | c | c | c | c | c | c | c |}
\hline
Concurrent Teams / SM & 16  & 15  & 14 & 13 & 12 & 8 & 4 & 2 & 1  \\\cline{1-10}
%----------------------------------------
Local and shared variables & 98 & 106 & 116 & 128 & 140 & 226 & 482 & 994 & 2018  \\\cline{1-10}
%----------------------------------------
Max registers / thread & 32 & 34 & 36 & 39 & 42 & 64 & 128 & 255 & 255  \\\cline{1-10}
%----------------------------------------
\end{tabular}
\end{center}
\caption{Maximum number of implicitly shared variables that can be allocated using all the available shared memory of a given SM of the K40 GPU if a certain number of concurrent teams per SM is enforced. For the number of registers per thread a thread limit of 128 threads was assumed.}
\label{tab:three}
\end{table*}
%--------------------------------------------------------------------------------

The shared memory footprint is computed as the sum of: (1) the shared stack size allocated in the PTX kernel prologue, the size of which is impacted by the number of both local and shared variables (2) the statically pre-allocated shared memory list of references to shared variables which is allocated in libomptarget, (3) the thread private state which is maintained in shared memory, the size of the thread private state is affected by the combination of OpenMP directives contained by the target region.

In the case of the one implicitly shared variable there are two local variables which are used for holding two arguments passed to the \texttt{\_\_kmpc\_kernel\_parallel} function call from the worker function: the team number and the thread number the function is called from.

The number of pre-allocated bytes should cover the requirements for cases used in practice. We go beyond the preallocated number of variables to test the impact on the shared memory of the device. We also test the dynamic allocation employed for the list containing the references to shared variables. The dynamic allocation of this buffer is performed by libomptarget.

\subsubsection{Sharing scalar variables}

The number of concurrent teams is affected by the number of registers required per thread and by the amount of shared memory required per team. We compute both these figures: the shared memory footprint as well as the number of registers allocated.

The results for the K40 and P100 GPUs are shown in Tables~\ref{tab:two} and~\ref{tab:twotwo} respectively. In the second column we include the shared memory footprint computed in Section~\ref{sec:footprint}. In the third column we show the size of the dynamically allocated global memory to hold the references to all the shared variables - note that variables are still held in shared memory.

The number of registers required by each thread is shown in column four. We compute the maximum number of concurrent teams by assuming 65536 registers per SM and 128 threads per team. We multiply the number of teams with the shared memory footprint of every team in order to obtain the total shared memory volume used on an SM basis.

\subsubsection{Sharing local arrays}

Similarly to the results in the previous section, Tables~\ref{tab:five} and~\ref{tab:six} measure the impact of array sharing on occupancy for the K40 and the P100 GPUs respectively.

\subsubsection{Shared memory impact on occupancy}
\label{sec:last}
We compute the number of variables that would be supported if register allocation was not a limiting factor. For each case we fix the number of concurrent teams. This implies a maximum number of registers for each thread and also a number of variables. We consider the team size to be 128 threads. In Table~\ref{tab:three}, for each fixed number of concurrent teams, we compute the number of variables required to use all available shared memory - 16KB - on the K40 GPU.

\subsection{Discussion}
\label{sec:discussion}

\subsubsection{K40 GPU}
In the results presented in this section we show that the footprint on shared memory for sharing scalar variables on a K40 GPU is relatively low. The number of registers allocated per thread is more of an occupancy limiter than shared memory is. The limits of our data sharing infrastructure are not reached even when a high number of concurrent teams is used since this is only possible for low number of shared variables.

The maximum shared memory footprint is achieved for 17 variables where the register usage of 42 limits the number of concurrent teams to 12 per SM, each team comprising 128 threads. The resulting shared memory footprint is 4332 bytes which is roughly 26\% of the available 16KB of shared memory on the K40 GPU in its default memory configuration.

In the experiment in Section~\ref{sec:last} we show that the maximum number of variables, either local or shared, that would reach the limit of the 16KB of shared memory on the K40 GPU, would need to lead to no more than 32 registers per thread. The likelihood of such a large number of variables being allocated within such a small number of registers leads to a very small subset of possible kernels.

The impact on occupancy of shared memory increases in the case of array variables. Register allocation in this case is low and the maximum number of concurrent teams is therefore limited by the 16KB of shared memory on the K40 default configuration. Handling a larger volume of shared data is possible by increasing the portion of shared memory of the GPU from 16KB to 32KB. Under this configuration the memory is no longer a bottleneck and the maximum number of concurrent teams can be achieved for up to five arrays.

The example in which only arrays are shared is an extreme example in the sense that no scalar variables are actually used. This means that the number of registers per thread is lower than in a practical hybrid example combining scalar and array variables.

\subsubsection{P100 GPU}
The P100 GPU benefits from an improved register allocation policy and a much larger shared memory area: 64 KB/SM. Our examples, due to their simplicity, do not benefit from the improved register allocation policy of the P100. The number of registers is roughly the same with that on the K40 GPU.

The analysis for sharing scalar variables is similar to that on the K40 GPU, this time the shared memory usage being even further from the device limit.

In the case of sharing array variables, the number of concurrent teams is not limited by the size of the shared memory despite the slightly lower register count.

For the P100 GPU, the maximum number of concurrent teams per SM is double that of the K40. Considering that the number of 32 bit registers per SM is the same, this leads to a maximum number of 16 registers per thread to achieve full concurrency. This leads to an even lower volume of shared memory per team.

\section{Limitations and further work}
\label{sec:limits}
The scheme described in this paper has several limitations which we plan to address in future work.

One solution open to consideration is to dynamically allocate shared memory instead of preallocating it. This would require the compiler to be augmented with a memory model to estimate the shared memory requirements of a given kernel. This would lead to a gain of an order of magnitude in performance for the cases in which we currently require more than the preallocated amount of shared variables.

There are currently two main design decisions that increase the amount of shared memory being allocated per team. These decisions have been taken to simplify the design of the scheme and are open to optimizations:
\begin{itemize}
\item The shared stack being allocated in the prologue of the PTX kernel is currently of the same size as the local stack. Any offsets previously computed for the local stack can just be re-used for the shared stack. This is increasing the shared memory footprint even when shared memory does not need to be used. We therefore would like to optimize the number of shared memory slots required by a given target region by developing an appropriate offset computation.
\item The detection of shared variables in the backend may end up including variables that do not need to be shared. Relying on Clang to mark the shared variables appropriately would be the more precise way of tracking which variables should be implicitly shared. This would also make it easier for the compiler to have a more precise estimate of the amount of shared memory required.
\end{itemize}

In this paper we discuss the sharing between master and worker threads which is a one-to-all pattern. OpenMP often requires an additional level of parallelism at worker level which may require the sharing of variables between all workers in an all-to-all pattern. This will significantly increase the shared memory volume required by OpenMP. For such cases in which the shared memory of the device is not enough to allow for the allocation of a shared memory stack, a runtime managed global memory stack needs to be employed instead. We aim to address this in future work.

\section{Conclusion}
\label{sec:conclusion}

In this paper we introduce a new data sharing scheme for implicitly sharing variables in OpenMP. The new scheme involves changes to the Clang code generation and the libomptarget runtime library. These changes rely on the NVPTX backend of LLVM to perform the lowering of variables to the shared memory of the device for cases required by OpenMP semantics. These changes are in line with the goal of generalizing the functionality of existing toolchains so that they can be used as targets by the device offloading capabilities of the OpenMP programming language in a maintainable way.

Despite the limitations of this scheme discussed in Section~\ref{sec:limits}, we show that the shared memory volume that it requires is relatively low for the well-established NVIDIA K40 GPU and even more so for the newer NVIDIA P100 GPU.

When all we share are scalar variables, the shared memory usage is no more than 4.3 KB. The actual bottleneck in most of these cases is shown to be the high register usage. For cases where register usage is low and the volume to be shared is high (for example, when sharing statically allocated local arrays), we show that even when running on a K40, given an appropriate shared memory to L1 ratio, reaching the limit of the shared memory can be avoided. There are of course cases for which the shared memory of the device is not enough. For such cases a global memory implementation of data sharing will be included in future work.

The experiments included in this paper, show that a model to estimate the shared memory needs of a kernel can be easily constructed for the benefit of both the users and the compiler. The relatively low shared memory footprint of this scheme ensures that, in practice, for user programs which contain a balanced number of shared array and scalar variables, the shared data can be fully contained by the shared memory of the device.

%\end{document}  % This is where a 'short' article might terminate

\begin{acks}
The authors would like to thank scientists at U.S. DOE laboratories for their valuable input and feedback on the development process of the OpenMP compiler. We would like to thank the LLVM community of reviewers for their help in getting our patches upstremed.

This paper is partially supported by the CORAL project LLNS Subcontract No. B604142.
\end{acks}

\bibliographystyle{ACM-Reference-Format}
\bibliography{sample-bibliography} 

\end{document}